\documentclass[a4paper]{article}
\usepackage{Odyssey2020}
\usepackage{epsfig,amssymb,amsmath}

\usepackage[nolist,nohyperlinks]{acronym}
\usepackage{multirow}
\usepackage[T1]{fontenc}
\usepackage{xcolor}
\usepackage{url}
\usepackage{caption}
\usepackage{soul}
\usepackage{hhline}
\usepackage{graphicx}

\newcommand{\SubItem}[1]{
    {\setlength\itemindent{15pt} \item[-] #1}
}

\acrodef{DFL}{Deep Feature Loss}
\acrodef{SV}{Speaker Verification}
\acrodef{SEGAN}{Speech Enhancement Generative Adversarial Network}
\acrodef{PLDA}{Probabilistic Linear Discriminant Analysis}
\acrodef{SOTA}{state-of-the-art}
\acrodef{SNR}{Signal-to-Noise Ratio}
\acrodef{SITW}{Speakers In The Wild}
\acrodef{PESQ}{Perceptual Evaluation of Speech Quality}
\acrodef{SDR}{Signal-to-Distortion Ratio}
\acrodef{WPE}{Weighted Prediction Error}
\acrodef{GAN}{Generative Adversarial Network}
\acrodef{ASR}{Automatic Speech Recognition}
\acrodef{CNN}{Convolutional Neural Network}
\acrodef{EER}{Equal Error Rate}
\acrodef{minDCF}{minimum Detection Cost Function}
\acrodef{WADA-SNR}{Waveform Amplitude Distribution Analysis for Signal-to-Noise Ratio}
\acrodef{LMFB}{log Mel-filterbank}
\acrodef{SA}{Signal Approximation}
\acrodef{CAN}{Context Aggregation Network}
\acrodef{TSE}{Temporal Squeeze Excitation}
\acrodef{TF}{Time-Frequency}
\acrodef{BN}{Batch Normalization}
\acrodef{LDE}{Learnable Dictionary Encoding}
\acrodef{E-TDNN}{Extended TDNN}
\acrodef{F-TDNN}{Factorized TDNN}
\acrodef{TDNN}{Time-Delay Neural Network}
\acrodef{WADA-SNR}{Waveform Amplitude Distribution Analysis}
\acrodef{AIR}{Aachen Impulse Response}
\acrodef{MVN}{Mean-Variance Normalized}
\acrodef{MFCC}{Mel-Frequency Cepstrum Coefficient}
\acrodef{LDA}{Linear Discriminant Analysis}
\acrodef{SE}{Squeeze Excitation}

\ninept     

\title{Analysis of Deep Feature Loss based Enhancement for Speaker Verification}

\name{Saurabh Kataria,
Phani Sankar Nidadavolu,
Jes\'us Villalba,
Najim Dehak}

\address{Center for Language and Speech Processing  \\
Johns Hopkins University, Baltimore, MD, USA \\
{\small \tt \{skatari1,snidada1,jvillal7,ndehak3\}@jhu.edu} }
\begin{document}
\maketitle

\begin{abstract}
\emph{Data augmentation} is conventionally used to inject robustness in Speaker Verification systems.
Several recently organized challenges focused on handling novel acoustic environments.
Deep learning based speech enhancement is a modern solution for this.
Recently, a study proposed to optimize the enhancement network in the activation space of a pre-trained auxiliary network.
This methodology, called \emph{deep feature loss}, greatly improved over the state-of-the-art conventional x-vector based system on a children speech dataset called \emph{BabyTrain}.
This work analyzes various facets of that approach and asks few novel questions in that context.
We first search for optimal number of auxiliary network activations, training data, and enhancement feature dimension.
Experiments reveal the importance of Signal-to-Noise Ratio filtering that we employ to create a large, clean, and naturalistic corpus for enhancement network training.
To counter the ``mismatch'' problem in enhancement, we find enhancing front-end (x-vector network) data helpful while harmful for the back-end (Probabilistic Linear Discriminant Analysis (PLDA)).
Importantly, we find enhanced signals contain complementary information to original.
Established by combining them in the front-end, this gives \textasciitilde40\% relative improvement over the baseline.
We also do an ablation study to remove a noise class from x-vector \emph{data augmentation} and, for such systems, we establish the utility of enhancement regardless of whether it has seen that noise class itself during training.
Finally, we design several dereverberation schemes to conclude ineffectiveness of \emph{deep feature loss} enhancement scheme for this task.
\end{abstract}

\section{Introduction}
Supervised deep learning based speech enhancement made significant progress in the last decade.
Notable works include masking~\cite{wang2018supervised} and mapping~\cite{xu2015regression} based approaches, \ac{SEGAN}~\cite{pascual2017segan}, \ac{DFL}~\cite{germain2018speech}, end-to-end metric optimization~\cite{kim2019end}, and a Transformer based approach~\cite{vaswani2017attention,kim2019transformer}.
Meanwhile, active research exists in the robustness of \ac{SV} systems~\cite{chung2019voxsrc,fan2019cn,kataria2019feature,shi2020robust}.
Another reason for interest in speech enhancement arises from the notion that it is considered as a modern solution to improve noise robustness in \ac{SV} systems~\cite{kataria2019feature,michelsanti2017conditional,shon2019voiceid}.
Such studies demonstrate that an explicit speech enhancement processing is beneficial to the \ac{SOTA} conventional x-vector and \ac{PLDA} based \ac{SV} system~\cite{villalba2019state}.
We refer to this methodology as \emph{task-specific} enhancement.
Prior work revealed its benefit for other tasks like Speaker Diarization~\cite{garcia2019speaker}, Language Recognition~\cite{frederiksen2018effectiveness}, and \ac{ASR}~\cite{ryanta2018enhancement}.

\begin{figure} 
    \includegraphics[width=0.35\textwidth]{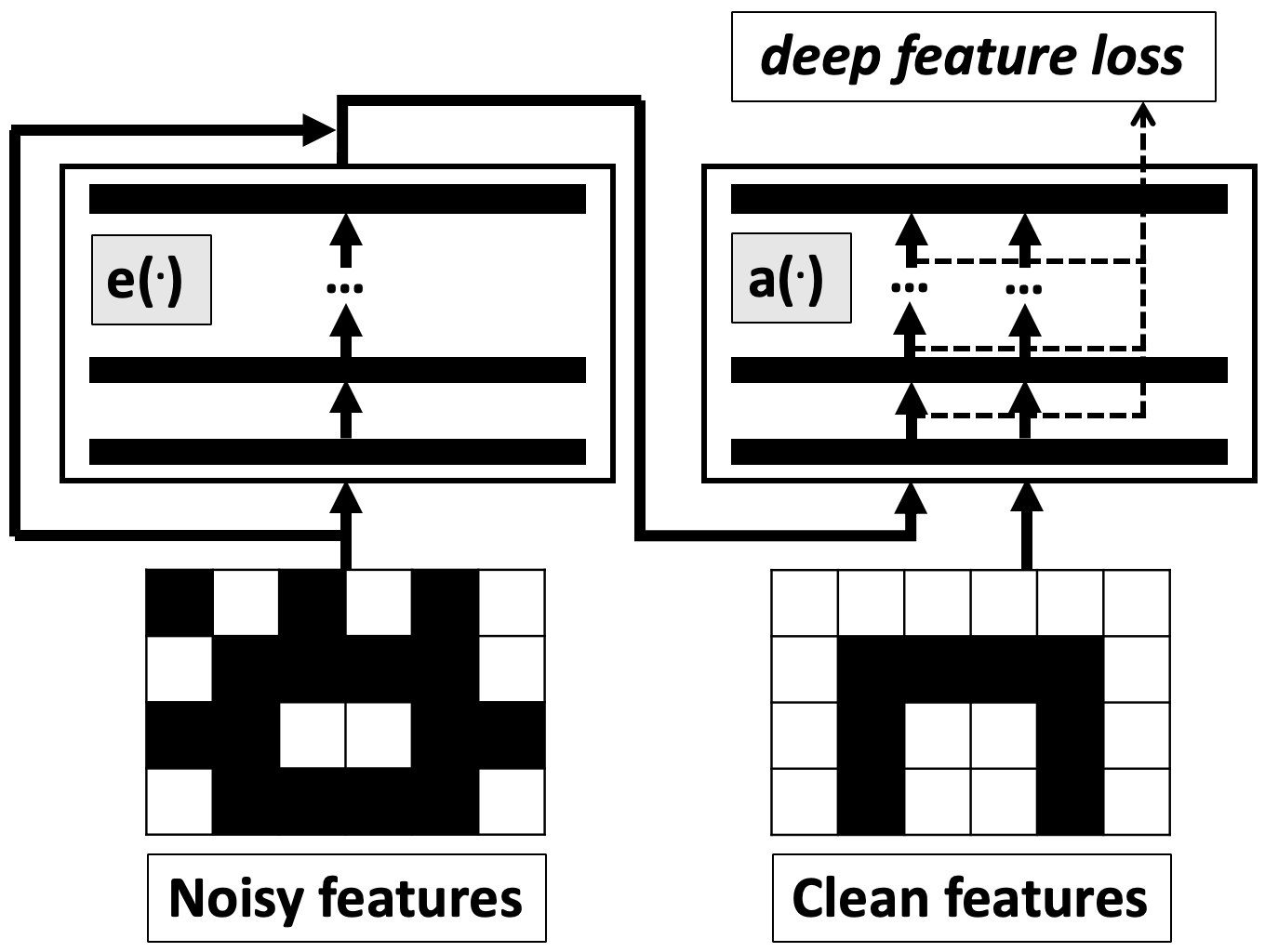}
    \caption{A schematic of \emph{deep feature loss} scheme}
    \label{fig:dfl}
    \vspace{-5mm}
\end{figure}

Building on \emph{perceptual loss}~\cite{johnson2016perceptual}, \cite{germain2018speech} proposed to learn speech enhancement using a pre-trained auxiliary network to obtain (deep feature) loss (Section \ref{sec:dfl}).
Authors observed that the usual supervised training with time-domain loss gives poor enhancement performance on low \ac{SNR} test signals, as confirmed with speech enhancement metrics like \ac{PESQ} and \ac{SDR}.
Therefore, they suggested to instead minimize the deviation of auxiliary network activations of enhanced and (reference) clean signals.
Here, enhanced signals refer to the output of the enhancement network (Figure \ref{fig:dfl}).

Recently, \cite{kataria2019feature} proposed a test-time feature denoising approach based on \cite{germain2018speech} and reported large gains over the \ac{SOTA} \emph{data augmented} x-vector based \ac{SV} system.
Since the conventional x-vector system can tackle clean signals such as in the \ac{SITW} dataset \cite{villalba2019state,mclaren2016speakers}, authors chose \ac{DFL} technique for its potential to handle low \ac{SNR} signals.
Due to their primary focus on final \ac{SV} performance, they chose the auxiliary network as speaker classification/embedding network.
Such enhancement preserves speaker information.
They reported results on a single-channel wide-band (16 KHz) dataset called \emph{BabyTrain}, which consists of daylong recordings of children speech in noisy and reverberant environments~\cite{vandam2016homebank}.
The main contribution of this study is to explore in-depth various facets of \ac{DFL}, ask some novel analysis-oriented questions, and present evaluation on real data (\emph{BabyTrain}).
This study is, therefore, similar in motivation to \cite{novotny2019analysis}.
We now describe the significance of all experiment sections.

Section \ref{sec:baseline} reproduces the gains observed with the \ac{DFL} based enhancement, as done in \cite{kataria2019feature}.
Furthermore, it judges the utility of activations from the deeper and, especially, the last layer (i.e. speaker embedding layer) of the auxiliary network.
Motivation for this comes from the common knowledge that a convolutional network contains high level information such as speaker identity primarily in the initial layers~\cite{dai2017very}.
\cite{germain2018speech} used only first few layers and our preliminary experiments on their setup revealed degradation by incorporating deeper layer activations.
However, their data setting was small (on VCTK corpus~\cite{yamagishi2019cstr}) and a much larger data setting such as ours is better suited to investigate this.

Section \ref{sec:traindatachoice} investigates the choice of training data for enhancement and auxiliary network.
For training enhancement network, it is imperative to have a clean, large, and naturalistic corpus.
For this, \cite{kataria2019feature} chose a (high) \ac{SNR}-filtered version of VoxCeleb~\cite{nagrani2020voxceleb,kim2008robust}.
In \ac{DFL} training, activations of noisy signals come from auxiliary network (Equation \ref{eq:dfl}).
Hence, it remains an open question if a stronger auxiliary network i.e. one trained with (noisy) \emph{data augmentations} is superior.
Training data choice is important to us because we focus on \emph{BabyTrain} and large ``in the wild'' public data releases such as \ac{SITW}~\cite{mclaren2016speakers}, VoxCeleb~\cite{nagrani2020voxceleb}, and CN-Celeb~\cite{fan2019cn} do not explicitly account for children speech.

Section \ref{sec:difffeat} asks whether it is beneficial to use higher dimensional features in the enhancement network.
For uniformity, we start with same features (40-dimensional \ac{LMFB}) for the enhancement, auxiliary, and x-vector network.
Then, we quantify the effect of increasing feature dimension for the former network while keeping it fixed for the others.
This idea of using different features for different networks is promising because most feature-domain enhancement studies work with spectrogram features.
They have higher dimension than the standard 40-D \ac{LMFB} features~\cite{villalba2019state} and we experiment with them too.

Section \ref{sec:beyond} explores whether enhancement of \ac{PLDA} and x-vector network data brings improvement on top of simple test set enhancement scheme.
Enhancement of data other than test sets can, potentially, counter the distortion introduced by enhancement and reduce the mismatch among test, \ac{PLDA}, and x-vector network data.
This is a notable problem in speech enhancement~\cite{ryanta2018enhancement,jahn2016wide,wang2019bridging}.
Note that enhancing x-vector data means training x-vector network with enhanced features.

Section \ref{sec:augwithenh} considers a different viewpoint to Section \ref{sec:beyond} and asks whether enhanced signals contain useful/complementary information to original signals.
We investigate this by including both enhanced and original signals in \ac{PLDA} and x-vector data.
Such analysis should provide insight into the nature of enhanced signals.
It is worthwhile to do as our enhancement setup is in (filterbank) feature domain and it is implausible to calculate time-domain metrics like \ac{SDR} and \ac{PESQ} for analysis.

Section \ref{sec:leaveoneout} tests the effectiveness of enhancement when a noise class is missing from \emph{data augmentation} of x-vector network.
While designing a generic x-vector based \ac{SV} system, it is a common practice to mix clean data with several noise classes such as music, babble, and general environmental noises.
We use this particular notion of \emph{data augmentation} in this study.
This may be not be optimal for the deployed environment and even cause performance degradation.
Thus, enhancement as a solution to robustness of \ac{SV} is attractive - provided the enhancer has good generalization property.
This section quantifies this generalization.
In this ``leave-one-out'' analysis, we, separately, consider the cases when enhancement has or has not seen the missing class.
This analysis is akin to finding harmful and/or superfluous noise class during \emph{data augmentation} and, thereby, similar in motivation to ablation and pruning work in deep learning~\cite{girshick2014rich,han2015deep}.

Section \ref{sec:dereverb} addresses an important extension to \cite{kataria2019feature}: effectiveness of \ac{DFL} enhancement for dereverberation for \ac{SV}.
\ac{WPE}~\cite{nakatani2008blind} is widely regarded as \ac{SOTA} dereverberation technique.
Recently, a \ac{GAN} based domain-adaptation work outperformed it in a large scale setting~\cite{nidadavolu2019low,nidadavolu2019unsupervised}.
We design several dereverberation schemes based on \ac{DFL}.
Several of such schemes combine denoising since dereverberation (alone) may be ineffective for final performance gains.

\section{Deep Feature Loss}
\label{sec:dfl}
Perceptual loss or Deep Feature Loss~\cite{johnson2016perceptual,germain2018speech} 
refers to the extraction of loss from a pre-trained auxiliary network by comparing its activations for enhanced and reference clean signal.
To obtain this, we manually pre-select few hidden layers of the auxiliary network.
Main idea is to enhance while retaining high level properties of signal.
This property depends on the choice of the auxiliary task.
With a speaker embedding/classification network (in our case), enhancement preserves speaker information.
Mathematically, \ac{DFL} using $j$ hidden layers of auxiliary network is:
\begin{equation}
    \begin{split}
        \mathcal{L}_{\text{DFL},[j]}(F_n,F_c) &= \sum_{i=1}^j \mathcal{L}_{\text{DFL},i}(F_n,F_c) \\
            &= \sum_{i=1}^j ||a_i(F_c) - a_i(e(F_n))||_{1,1}.
    \end{split}
\label{eq:dfl}
\end{equation}
Also,
\begin{equation}
    \mathcal{L}_{\text{DFL,emb}} = \mathcal{L}_{\text{DFL},[5]} + \mathcal{L}_{\text{emb}}.
\end{equation}
Here, $F_n$ and $F_c$ refers to noisy and clean feature matrices of size $D \times T$, $D$ is the feature dimension, $T$ is number of frames, $j$ is the number of hidden layers considered for \ac{DFL} computation, $i$ is the index for such layers, $a(\cdot)$ is the auxiliary network, $e(\cdot)$ is the enhancement network.
``1,1'' is $l_1$ loss for matrix.
A corresponding visual description is in Figure \ref{fig:dfl}.
The maximum value of $j$ is $L=5$.
They refer to 5 equidistant hidden layers preselected in our auxiliary network.
We handle final layer activations exclusively by the loss denoted by $\mathcal{L}_{\text{DFL,emb}}$.
$\mathcal{L}_{\text{FL}}$ refers to the usual \emph{feature loss} i.e. without using auxiliary network.
Importantly, we do not use x-vector network itself for extracting \ac{DFL} because it may be not be optimal as noted in Section \ref{sec:traindatachoice}.

\section{Neural Networks Architectures}
\subsection{Enhancement network}
We choose \ac{CNN} based \ac{CAN} from \cite{kataria2019feature} except with higher number of channels (90).
It is inspired by \ac{CAN} in \cite{germain2018speech}.
Its main features are linearly increasing dilations (1 to 8), eight convolution layers, Adaptive \ac{BN}, LeakyReLU activations, and three \ac{TSE}~\cite{kataria2019feature} connections along with residual connections.

Final layer linearly maps the output to input dimension and a subsequent logarithm operation predicts the \ac{TF} mask~\cite{wang2018supervised}.
To mimic \ac{SA} loss~\cite{wang2018supervised}, we add this log-domain mask to the original input (multiplication in linear domain) to predict the final enhanced features.
We found this global skip connection significantly helpful in our preliminary experiments.
The network has a context length of 73 frames and 10.2M number of parameters.
Since the main feature of \ac{CAN} is high context, we tried increasing its receptive field but observed degradation in our preliminary experiments.

\subsection{Auxiliary network}
The auxiliary network used in this work is the 16KHz version ResNet-34 network described in~\cite{villalba2018jhu,snyder2019jhu,villalba2019state}.
We select this network due to its good performance on \ac{SV}~\cite{villalba2018jhu}.
It is a 2D CNN based ResNet-34 residual network~\cite{he2016deep} with \ac{LDE} pooling~\cite{cai2018novel} and Angular Softmax loss function~\cite{liu2017sphereface,liu2016large}.
The dictionary size of \ac{LDE} is 64 and the network has 5.9M parameters.

\subsection{x-vector network}
We choose \ac{E-TDNN} introduced in \cite{snyder2019speaker}.
\ac{E-TDNN} greatly improves upon \ac{TDNN} by interleaving dense layers with convolution layers and employing a (slightly) wider temporal context.
Total trainable parameters are 10M.
A summary of its exact specification is in~\cite{villalba2019state}.
\cite{kataria2019feature} prefers a larger \ac{F-TDNN} network due to its superior performance than \ac{E-TDNN}.
Since several of our experiments require re-training of the x-vector network, we choose \ac{E-TDNN} to facilitate faster experimentation.
Note that \ac{E-TDNN} gives competitive performance~\cite{villalba2019state} and, therefore, is suitable for our analysis-oriented work.

\section{Experimental Setup}
\subsection{Dataset details}
We combine VoxCeleb1 and VoxCeleb2 \cite{chung2018voxceleb2,nagrani2020voxceleb,nagrani2017voxceleb} to create \textit{voxceleb}.
We, then, concatenate utterances from the same video to create \textit{voxcelebcat} (or \textit{vc}).
This gives us 2710 hrs of relatively clean audio with 7185 speakers.
\textit{voxcelebcat\_div2} (or \textit{vc\_div2}) refers to a random 50\% subset of \textit{voxcelebcat}.
We use a SNR estimation algorithm called \ac{WADA-SNR} \cite{kim2008robust} to retain top 50\% clean samples from \textit{voxcelebcat} to create \textit{voxcelebcat\_wadasnr} (or \textit{vc.w}).
This is 1665 hrs of audio with 7104 speakers.

To create noisy counterpart, we use noise utterances from MUSAN~\cite{snyder2015musan} and DEMAND~\cite{thiemann2013diverse} corpora.
We make the reverberant counterpart using impulse responses of small and medium size rooms from the \ac{AIR} database~\cite{jeub2009binaural}.
A 90-10 split gives us the training and validation lists for the enhancement system.
Lastly, using noise files, we corrupt \textit{voxcelebcat} to form \textit{voxcelebcat\_combined} (\textit{vcc}).
Its size is three times as that of \textit{voxcelebcat}.
``libri'' refers to LibriSpeech corpus~\cite{panayotov2015librispeech}.
Unless specified otherwise, we train the auxiliary network and x-vector network with \textit{voxcelebcat\_wadasnr} and \textit{voxcelebcat\_combined} respectively.

For evaluation on real data, we choose \emph{BabyTrain} corpus which is based on the Homebank repository~\cite{vandam2016homebank}.
It consists of day-long children speech in uncontrolled noisy and reverberant environments.
Recordings are in the presence of several (dynamic) number of background speakers.
Training data for diarization and detection (\textit{adaptation data}) has duration of 130 and 120 hrs respectively.
For evaluation, enrollment is 95hrs and has 595 speakers, while test data is 30 hrs with 158 speakers.
The classification of enrollment and test utterances is as follows.
\textit{test>=$n$} and \textit{enroll=$m$} refers to test and enrollment utterances of minimum $n$ and equal to $m$ seconds from the speaker of interest respectively with $n \in \{0,5,15,30\}$ and $m \in \{5,15,30\}$.
For enrollment utterances, time marks of the target speaker are present but not for the test utterances.
There may be multiple speakers present in the test utterances.
Data split description and respective scripts were devised in JSALT 2020 workshop and are available online\footnote{\url{https://github.com/jsalt2019-diadet}}.

\subsection{Training details}
We train enhancement network with batch size of 32, learning rate of 0.001 (exponentially decreasing), 6 epochs, Adam optimizer~\cite{kingma2014adam}, and 500 frames (5s audio).
Its code is available online as ``DFL\_TSEResCAN2d\_SmallContext\_LogSigMask\_BNIn''\footnote{\url{https://github.com/jsalt2019-diadet/jsalt2019-diadet/blob/master/egs/sitw_noisy/v1.pyfb/steps_pyfe/enh_models/models.py}}.
Unless otherwise stated, input features are un-normalized 40-D \ac{LMFB} features.
We train the auxiliary network with batch size of 128, number of epochs as 50, optimizer as Adam~\cite{kingma2014adam}, learning rate of 0.0075 (exponentially decreasing) with warmup~\cite{vaswani2017attention}, and sequences of 800 frames (8s audio).
Since this network is a \ac{CNN}, we use mean-normalized \ac{LMFB} features which have spatial information contrary to \ac{MFCC} features.
To account for this normalization mismatch with the enhancement network, we insert an online mean normalization between them during \ac{DFL} training.
For x-vector network training, we use Kaldi~\cite{povey2011kaldi} scripts with 40-D \ac{LMFB} features which have silence removed and are mean-normalized.

\subsection{Evaluation details}
\label{sec:evaldet}
The \ac{PLDA}-based back-end consists of a 200-D \ac{LDA} with generative Gaussian SPLDA~\cite{villalba2018jhu}.
Additionally, we use a diarization system since \textit{BabyTrain} consists of \textit{babble} noise (background speakers).
For this, we followed the Kaldi x-vector Callhome diarization recipe~\cite{snyder2018xvector}.
Details are in the \textit{JHU-CLSP diarization system} as described in~\cite{villalba2018jhu}.
Note that, in general, ``enhancement of test set'' refers to enhancing test, enroll, and \textit{adaptation data}.
For the final evaluation, we use standard metrics like \ac{EER} and \ac{minDCF} at target prior $p=0.05$ (NIST SRE18 VAST operating point~\cite{sadjadi20172016}).
Except Kaldi based x-vector training, we develop all framework using Hyperion library\footnote{\url{https://github.com/jsalt2019-diadet/hyperion}} and Pytorch~\cite{paszke2019pytorch}.

\begin{table} 
    \centering
    \caption{Baseline results}
    \resizebox{0.47\textwidth}{!}{
        \begin{tabular}{|c|cccc|c|}
            \hline
            \textbf{EER} & \textbf{test>=30s} & \textbf{test>=15s} & \textbf{test>=5s} & \textbf{test>=0s} & \textbf{mean} \\
                \hline
                no-enh & 5.78 & 8.78 & 12.34 & 12.71 & 9.90\\
                \hline
                $\mathcal{L}_{\text{DFL},[5]}$ (*) & \textbf{5.14} & \textbf{7.17} & 11.02 & 11.41 & \textbf{8.68}\\
                \hline
                $\mathcal{L}_{\text{FL}}$ & 6.28 & 8.90 & 12.35 & 12.71 & 10.06\\
                \hline
                $\mathcal{L}_{\text{DFL},[5]} + \mathcal{L}_{\text{FL}}$ & 5.66 & 8.11 & 11.40 & 11.79 & 9.24\\
                \hline
                $\mathcal{L}_{\text{DFL},[5]} + \mathcal{L}_{\text{emb}}$ & 5.38 & 7.84 & 11.07 & 11.47 & 8.94\\
                \hline
                $\mathcal{L}_{\text{DFL},[4]}$ & 5.63 & 7.96 & 11.26 & 11.62 & 9.12\\
                \hline
                $\mathcal{L}_{\text{DFL},[3]}$ & 5.32 & 7.75 & \textbf{10.83} & \textbf{11.18} & 8.77\\
                \hline
                $\mathcal{L}_{\text{DFL},[2]}$ & 5.93 & 8.36 & 11.79 & 12.16 & 9.56\\
                \hline
                $\mathcal{L}_{\text{DFL},[1]}$ & 5.73 & 8.38 & 11.84 & 12.19 & 9.54\\
                \hline
                \multicolumn{6}{|c|}{}\\
                \hline
                \textbf{minDCF} & \textbf{test>=30s} & \textbf{test>=15s} & \textbf{test>=5s} & \textbf{test>=0s} & \textbf{mean} \\
                \hline
                no-enh & 0.255 & 0.386 & 0.492 & 0.499 & 0.408\\
                \hline
                $\mathcal{L}_{\text{DFL},[5]}$ (*) & \textbf{0.204} & 0.333 & 0.441 & 0.448 & \textbf{0.357}\\
                \hline
                $\mathcal{L}_{\text{FL}}$ & 0.239 & 0.370 & 0.478 & 0.485 & 0.393\\
                \hline
                $\mathcal{L}_{\text{DFL},[5]} + \mathcal{L}_{\text{FL}}$ & 0.218 & 0.343 & 0.452 & 0.459 & 0.368\\
                \hline
                $\mathcal{L}_{\text{DFL},[5]} + \mathcal{L}_{\text{emb}}$ & 0.210 & \textbf{0.331} & \textbf{0.439} & \textbf{0.447} & \textbf{0.357}\\
                \hline
                $\mathcal{L}_{\text{DFL},[4]}$ & 0.213 & 0.342 & 0.452 & 0.459 & 0.367\\
                \hline
                $\mathcal{L}_{\text{DFL},[3]}$ & 0.215 & 0.334 & 0.441 & 0.449 & 0.360\\
                \hline
                $\mathcal{L}_{\text{DFL},[2]}$ & 0.218 & 0.338 & 0.446 & 0.453 & 0.364\\
                \hline
                $\mathcal{L}_{\text{DFL},[1]}$ & 0.215 & 0.334 & 0.441 & 0.448 & 0.360\\
                \hline
        \end{tabular}
    }
    \label{tab:baseline}
    \vspace{-5mm}
\end{table}

\section{Experiments}
\subsection{Baseline results}
\label{sec:baseline}
In Table \ref{tab:baseline}, we reproduce the claims of \cite{kataria2019feature}.
Last column refers to the mean metric value per row.
We organize results for \ac{EER} and \ac{minDCF} separately.
Boldface result signify the best value achieved per column per metric. 
Note that x-vector network is trained with augmentation in all cases and enhancement is applied on \textit{adaptation data}, enrollment, and test utterances.
That is, we use the default test-time enhancement scheme as mentioned in Section \ref{sec:evaldet}.

``no-enh'' refers to the case when enhancement is not used in the \ac{SV} pipeline. 
$\mathcal{L}_{\text{DFL},[5]}$ refers to the results obtained with \ac{DFL} using all $L=5$ intermediate hidden layers of the auxiliary network.
We note relative improvement of 12.3\% and 12.5\% for \ac{EER} and \ac{minDCF} respectively w.r.t. ``no-enh''.
Feature loss leads to lesser gains contrary to degradation caused in \cite{kataria2019feature}.
This variation is perhaps due to use of a different x-vector network in this work.
Combining it with \ac{DFL} gives better results.
We note that adding auxiliary network speaker embedding layer loss 
$(\mathcal{L}_{\text{DFL},[5]} + \mathcal{L}_{\text{emb}})$ does not lead to improvement.
This suggests that all hidden activations from auxiliary network need not be useful for final performance.
Using lesser number of layers in \ac{DFL} does not lead to consistent observation.
Nevertheless, $\mathcal{L}_{\text{DFL},[5]} (*)$ gives best performance for both metrics and it serves as the baseline for this work.
These baseline results are present in all results tables under different names but all denoted by (*).

Importantly, note that results under ``test>=0s'' represent final average performance on \emph{BabyTrain}.
``mean'' refers to the weighted mean performance with higher weight for longer test trials.
In practice, it is uncommon to have very small test utterances.
Therefore, for this practical significance, we consider ``mean'' for final model comparisons in this work.
For simplicity in reading all tables, reader may focus on ``mean'' performance.

\begin{table} 
    \centering
    \caption{Choice of training data for enhancement and auxiliary network. ``vc'' is VoxCeleb, ``vc.w'' is 50\% WADASNR-filtered VoxCeleb, ``vc\_div2'' is 50\% random subsampled VoxCeleb, ``vcc'' is VoxCeleb with 3x augmentations, ``libri'' is LibriSpeech.}
    \resizebox{0.47\textwidth}{!}{
        \begin{tabular}{|c|cccc|c|}
            \hline
            \textbf{EER} & \textbf{test>=30s} & \textbf{test>=15s} & \textbf{test>=5s} & \textbf{test>=0s} & \textbf{mean} \\
            \hline
                no-enh & 5.78 & 8.78 & 12.34 & 12.71 & 9.90\\
                \hline
                vc.w-vc.w (*) & \textbf{5.14} & \textbf{7.17} & \textbf{11.02} & 11.41 & \textbf{8.68}\\
                \hline
                vc.w-vc & 5.63 & 8.12 & 11.37 & 11.74 & 9.22\\
                \hline
                vc.w-vcc & 5.19 & 7.81 & \textbf{11.02} & \textbf{11.39} & 8.85\\
                \hline
                vc-vc.w & 5.33 & 7.87 & 11.17 & 11.57 & 8.99\\
                \hline
                vc-vc & 5.62 & 8.25 & 11.63 & 12.00 & 9.38\\
                \hline
                vc-vcc & 5.43 & 8.16 & 11.44 & 11.80 & 9.21\\
                \hline
                vc\_div2-vc.w & 5.29 & 8.10 & 11.51 & 11.89 & 9.20\\
                \hline
                libri-vc.w & 6.00 & 9.08 & 12.68 & 13.06 & 10.21\\
                \hline
            \multicolumn{6}{|c|}{}\\
            \hline
            \textbf{minDCF} & \textbf{test>=30s} & \textbf{test>=15s} & \textbf{test>=5s} & \textbf{test>=0s} & \textbf{mean} \\
            \hline
                no-enh & 0.255 & 0.386 & 0.492 & 0.499 & 0.408\\
                \hline
                vc.w-vc.w (*) & \textbf{0.204} & 0.333 & 0.441 & 0.448 & \textbf{0.357}\\
                \hline
                vc.w-vc & 0.215 & 0.335 & 0.444 & 0.451 & 0.361\\
                \hline
                vc.w-vcc & 0.210 & \textbf{0.330} & \textbf{0.440} & \textbf{0.447} & \textbf{0.357}\\
                \hline
                vc-vc.w & 0.220 & 0.344 & 0.450 & 0.457 & 0.368\\
                \hline
                vc-vc & 0.226 & 0.345 & 0.453 & 0.460 & 0.371\\
                \hline
                vc-vcc & 0.222 & 0.349 & 0.456 & 0.463 & 0.373\\
                \hline
                vc\_div2-vc.w & \textbf{0.204} & 0.335 & 0.444 & 0.451 & 0.359\\
                \hline
                libri-vc.w & 0.232 & 0.357 & 0.464 & 0.471 & 0.381\\
                \hline
        \end{tabular}
    }
    \label{tab:traindatachoice}
     \vspace{-5mm}
\end{table}

\subsection{Choice of training data for enhancement and auxiliary network}
\label{sec:traindatachoice}
Table \ref{tab:traindatachoice} presents the results obtained with different choice of training data for enhancement and auxiliary network.
Here, training data for enhancement network refers to the clean data counterpart required for creating training pairs for supervised learning.
A preliminary \ac{WADA-SNR} analysis of VoxCeleb (``vc'') revealed the presence of several low \ac{SNR} signals.
For this reason, we use \ac{SNR} estimation to retain top 50\% clean utterances from ``vc'' to form ``vc.w''.
The second column of Table \ref{tab:traindatachoice} specifies the training data for enhancement and auxiliary network (separated by ``-'') respectively.

We make few prominent observations.
First, by comparing enhancers trained with ``vc'' and ``vc.w'' as enhancement network training data, we find using full VoxCeleb (``vc'') harmful for both metrics.
This suggests ``vc'' may not be clean enough for training enhancer and some filtering may be necessary.
Second, using ``vc\_div2'' in place of ``vc.w'' degrades \ac{EER}, which suggests a SNR-based filtering is better than random subsampling.
Third, to test the hypothesis that a cleaner data (LibriSpeech) helps further, we find that it gives worst performance.
This establishes the superiority of VoxCeleb, perhaps, due to its diverse and spontaneous conversation nature, which is contrary to the read speech nature of LibriSpeech.
Fourth, in our \ac{DFL} formulation, we obtain activations of noisy samples from the auxiliary network (Equation \ref{eq:dfl}).
We do not observe gains by using a stronger auxiliary network (trained with ``vc'' or ``vcc'').
This is contrary to the popular notion that even clean test files benefit from data augmentation~\cite{snyder2018x}.
This indicates that using x-vector network for \emph{deep feature loss} extraction may not be optimal, as hinted in Section \ref{sec:dfl}.
To sum up, we obtain best results with SNR-filtered VoxCeleb for both networks (``vc.w-vc.w'').

\begin{table} 
    \centering
    \caption{Enhancement with mismatch between enhancement and x-vector/aux. network acoustic features. 
    First column indicates enhanced features, x-vec/aux. networks always use 40D LMFB.}
    \resizebox{0.47\textwidth}{!}{
        \begin{tabular}{|c|cccc|c|}
        \hline
        \textbf{EER} & \textbf{test>=30s} & \textbf{test>=15s} & \textbf{test>=5s} & \textbf{test>=0s} & \textbf{mean} \\
        \hline
        no-enh & 5.78 & 8.78 & 12.34 & 12.71 & 9.90\\
        \hline
        LMFB-40D (*) & 5.14 & 7.17 & 11.02 & 11.41 & \textbf{8.69}\\
        \hline
        LMFB-80D & 6.46 & 10.14 & 13.83 & 14.22 & 11.16\\
        \hline
        LMFB-100D & 6.43 & 9.76 & 13.40 & 13.79 & 10.85\\
        \hline
        LMFB-120D & 6.84 & 10.14 & 13.77 & 14.17 & 11.23\\
        \hline
        spectrogram-256D & 5.72 & 8.91 & 12.49 & 12.84 & 9.99\\
        \hline
        \multicolumn{6}{|c|}{}\\
        \hline
        \textbf{minDCF} & \textbf{test>=30s} & \textbf{test>=15s} & \textbf{test>=5s} & \textbf{test>=0s} & \textbf{mean} \\
        \hline
        no-enh & 0.255 & 0.386 & 0.492 & 0.499 & 0.408\\
        \hline
        LMFB-40D (*) & 0.204 & 0.333 & 0.441 & 0.448 & \textbf{0.357}\\
        \hline
        LMFB-80D & 0.276 & 0.444 & 0.546 & 0.553 & 0.455\\
        \hline
        LMFB-100D & 0.284 & 0.436 & 0.539 & 0.545 & 0.451\\
        \hline
        LMFB-120D & 0.288 & 0.446 & 0.546 & 0.552 & 0.458\\
        \hline
        spectrogram-256D & 0.242 & 0.390 & 0.492 & 0.498 & 0.406\\
        \hline
        \end{tabular}
    }
    \label{tab:difffeat}
     \vspace{-5mm}
\end{table}

\subsection{Enhancement with mismatch between enhancement and x-vector/aux. network acoustic features}
\label{sec:difffeat}
Table \ref{tab:difffeat} presents the results by varying the feature used in the enhancement network.
Result rows specify the feature dimension against the name of the feature.
Features (40-D \ac{LMFB}) for the auxiliary and x-vector network remain unchanged.
A trainable linear layer bridges enhancement and auxiliary network to handle the mismatch of the feature dimensions for these networks.
This bridge comes before the global skip connection of the enhancement network.
We note that all higher dimensional feature models result in similar level of degradation except for spectrogram which leads to lesser degradation.
As an additional evidence, we observed higher variance in the training and validation losses for these networks.
This degradation is perhaps because learning with higher dimensional features require more data.
A fair comparison study should, correspondingly, vary the training data amounts but we do not investigate that.
Another option to avoid degradation could be to use same higher-dimensional features for all three networks.
However, that leads to increased training complexity and, possibly, worse performance as apparent by the popularity of low-dimensional features like 40-D \ac{LMFB} in \ac{SOTA} \ac{SV} systems~\cite{villalba2019state}.

\begin{table} 
    \centering
    \caption{Effect of enhancing PLDA and/or x-vector data on top of test set enhancement}
    \resizebox{0.47\textwidth}{!}{
        \begin{tabular}{|c|cccc|c|}
            \hline
            \textbf{EER} & \textbf{test>=30s} & \textbf{test>=15s} & \textbf{test>=5s} & \textbf{test>=0s} & \textbf{mean} \\
            \hline
                no-enh & 5.78 & 8.78 & 12.34 & 12.71 & 9.90\\
                \hline
                test (*) & 5.14 & \textbf{7.17} & 11.02 & 11.41 & \textbf{8.68}\\
                \hline
                PLDA,test & \textbf{4.93} & 7.58 & \textbf{10.93} & \textbf{11.34} & 8.70\\
                \hline
                train,test & 5.36 & 8.01 & 11.25 & 11.63 & 9.06\\
                \hline
                train,PLDA,test & 6.74 & 10.23 & 14.27 & 14.71 & 11.49\\
            \hline
            \multicolumn{6}{|c|}{}\\
            \hline
            \textbf{minDCF} & \textbf{test>=30s} & \textbf{test>=15s} & \textbf{test>=5s} & \textbf{test>=0s} & \textbf{mean} \\
            \hline
                no-enh & 0.255 & 0.386 & 0.492 & 0.499 & 0.408\\
                \hline
                test (*) & 0.204 & 0.333 & 0.441 & 0.448 & 0.357\\
                \hline
                PLDA,test & 0.211 & 0.340 & 0.449 & 0.456 & 0.364\\
                \hline
                train,test & \textbf{0.199} & \textbf{0.315} & \textbf{0.425} & \textbf{0.432} & \textbf{0.343}\\
                \hline
                train,PLDA,test & 0.295 & 0.443 & 0.551 & 0.558 & 0.462\\
                \hline
        \end{tabular}
    }
    \label{tab:beyond}
    \vspace{-5mm}
\end{table}
\vspace{-2mm}
\subsection{Effect of enhancing PLDA and/or x-vector data on top of test set enhancement}
\label{sec:beyond}
Table \ref{tab:beyond} presents the results for systems with enhancement of \ac{PLDA} and/or x-vector train data (``train'') on top of test, enroll, \emph{adaptation data} enhancement (``test'').
First column lists the datasets that undergo enhancement processing.
We find enhancing \ac{PLDA} data (slightly) harmful.
Enhancing x-vector data gives best minDCF, while enhancing x-vector and \ac{PLDA} data gives worst performance, even worse than the case of no enhancement.
This suggests that \ac{PLDA} is susceptible to enhancement processing.
This finding is contrary to the notion that enhancement of all datasets solve the mismatch problem~\cite{wang2019bridging}.

\vspace{-2mm}
\subsection{Augmentation with enhanced features}
\label{sec:augwithenh}
In Table \ref{tab:augwithenh}, ``test (*)'' and ``PLDA,test'' (from Table \ref{tab:beyond}) represent enhancement of test set and test set along with \ac{PLDA} data respectively.
To gain insight into the nature of enhanced signals, we investigate if they contain complementary information to original signals.
``aug-in-PLDA'' refers to including enhanced signals with original (non-enhanced) in \ac{PLDA} data.
In Section \ref{sec:beyond}, we noted that training \ac{PLDA} with enhanced data gives worse performance compared to training with original data.
Here, combining them causes further degradation.

The next experiment is ``aug-in-train'', 
which refers to training x-vector data with original as well as enhanced data.
This doubles the training data and time but, nevertheless, counts for a fair investigation since we train all x-vector networks till convergence and don't introduce any new data here.
Note that we assign same speaker label to enhanced signal as the original.
Doing this bring huge (relative) improvements of \textasciitilde40\% in both metrics.
This strongly establishes our hypothesis that enhanced signals contain useful complementary information.
This is a novel finding albeit computationally expensive.
``aug-in-train,PLDA'' is an extension of ``aug-in-train''.
It refers to inclusion of enhanced and original signals in x-vector as well as \ac{PLDA} data.
This leads to some degradation with respect to ``aug-in-train''.
Thus, it is our consistent observation that \ac{PLDA} is susceptible to enhancement processing and it is best trained with unenhanced data.
It is useful to reiterate that in our enhancement schemes, test set is always enhanced.

\begin{table}[htbp]
    \centering
    \caption{Augmentation with enhanced features}
    \resizebox{0.47\textwidth}{!}{
        \begin{tabular}{|c|cccc|c|}
            \hline
            \textbf{EER} & \textbf{test>=30s} & \textbf{test>=15s} & \textbf{test>=5s} & \textbf{test>=0s} & \textbf{mean} \\
            \hline
                no-enh & 5.78 & 8.78 & 12.34 & 12.71 & 9.90\\
                \hline
                test (*) & 5.14 & 7.17 & 11.02 & 11.41 & 8.68\\
                \hline
                PLDA,test & 4.93 & 7.58 & 10.93 & 11.34 & 8.70\\
                \hline
                aug-in-PLDA & 5.31 & 8.06 & 11.48 & 11.87 & 9.18\\
                \hline
                aug-in-train & \textbf{3.34} & \textbf{4.99} & \textbf{7.53} & \textbf{7.92} & \textbf{5.95}\\
                \hline
                aug-in-train,PLDA & 3.38 & 5.13 & 7.78 & 8.19 & 6.12\\
                \hline
            \multicolumn{6}{|c|}{}\\
            \hline
            \textbf{minDCF} & \textbf{test>=30s} & \textbf{test>=15s} & \textbf{test>=5s} & \textbf{test>=0s} & \textbf{mean} \\
            \hline
                no-enh & 0.255 & 0.386 & 0.492 & 0.499 & 0.408\\
                \hline
                test (*) & 0.204 & 0.333 & 0.441 & 0.448 & 0.357\\
                \hline
                PLDA,test & 0.211 & 0.340 & 0.449 & 0.456 & 0.364\\
                \hline
                aug-in-PLDA & 0.219 & 0.350 & 0.459 & 0.466 & 0.374\\
                \hline
                aug-in-train & \textbf{0.128} & \textbf{0.209} & \textbf{0.300} & \textbf{0.309} & \textbf{0.237}\\
                \hline
                aug-in-train,PLDA & 0.132 & 0.215 & 0.307 & 0.315 & 0.242\\
                \hline
        \end{tabular}
    }
    \label{tab:augwithenh}
\end{table}

\begin{table}[!h]
    \centering
    \caption{Leave-one-out noise class in x-vector data. Each block leaves one noise type from x-vector training. The first row in each block
    is without enhancement, ``enh-unseen'' trains enh. without the left-out noise, ``enh-seen'' trains enh. with all noises.}
    \resizebox{0.47\textwidth}{!}{
        \begin{tabular}{|c|cccc|c|}
            \hline
            \textbf{EER} & \textbf{test>=30s} & \textbf{test>=15s} & \textbf{test>=5s} & \textbf{test>=0s} & \textbf{mean} \\
            \hline
                no-enh & 5.78 & 8.78 & 12.34 & 12.71 & 9.90\\
                test-enh (*) & 5.14 & 7.17 & 11.02 & 11.41 & \underline{8.68}\\
                \hline
                \textit{noise} & 7.36 & 10.90 & 15.02 & 15.44 & 12.18\\
                enh-unseen & 5.88 & 9.59 & 13.51 & 13.93 & \underline{10.73}\\
                enh-seen & 6.30 & 9.87 & 13.97 & 14.38 & 11.13\\
                \hline
                \textit{music} & 4.99 & 7.01 & 9.93 & 10.28 & 8.05\\
                enh-unseen & 4.42 & 6.52 & 9.54 & 9.96 & 7.61\\
                enh-seen & 4.35 & \textbf{6.38} & \textbf{9.34} & \textbf{9.74} & \underline{\textbf{7.45}}\\
                \hline
                \textit{babble} & 4.98 & 7.59 & 11.04 & 11.46 & 8.77\\
                enh-unseen & 4.13 & 6.56 & 9.61 & 10.03 & \underline{7.58}\\
                enh-seen & \textbf{4.07} & 6.64 & 9.82 & 10.26 & 7.70\\
                \hline
                \textit{chime3bg} & 5.49 & 7.66 & 10.69 & 11.04 & 8.72\\
                enh-unseen & 4.83 & 7.48 & 10.51 & 10.88 & \underline{8.43}\\
                enh-seen & 4.97 & 7.66 & 10.70 & 11.05 & 8.59\\
            \hline
            \multicolumn{6}{|c|}{}\\
            \hline
            \textbf{minDCF} & \textbf{test>=30s} & \textbf{test>=15s} & \textbf{test>=5s} & \textbf{test>=0s} & \textbf{mean} \\
            \hline
                no-enh & 0.255 & 0.386 & 0.492 & 0.499 & 0.408\\
                test-enh (*) & \textbf{0.204} & 0.333 & 0.441 & 0.448 & 0.357\\
                \hline
                \textit{noise} & 0.414 & 0.525 & 0.618 & 0.624 & 0.545\\
                enh-unseen & 0.334 & 0.474 & 0.572 & 0.578 & \underline{0.489}\\
                enh-seen & 0.333 & 0.484 & 0.586 & 0.592 & 0.499\\
                \hline
                \textit{music} & 0.255 & 0.355 & 0.454 & 0.461 & 0.381\\
                enh-unseen & 0.217 & 0.327 & 0.424 & 0.432 & 0.350\\
                enh-seen & 0.213 & 0.326 & 0.425 & 0.433 & \underline{0.349}\\
                \hline
                \textit{babble} & 0.247 & 0.357 & 0.458 & 0.465 & 0.382\\
                enh-unseen & 0.213 & 0.324 & 0.423 & 0.431 & 0.348\\
                enh-seen & 0.206 & \textbf{0.320} & \textbf{0.419} & \textbf{0.426} & \underline{\textbf{0.343}}\\
                \hline
                \textit{chime3bg} & 0.302 & 0.420 & 0.523 & 0.530 & 0.444\\
                enh-unseen & 0.264 & 0.402 & 0.509 & 0.515 & 0.423\\
                enh-seen & 0.257 & 0.392 & 0.499 & 0.506 & \underline{0.414}\\
                \hline
        \end{tabular}
    }
    \label{tab:leaveoneout}
    \vspace{-5mm}
\end{table}

\subsection{Leave-one-out noise class in x-vector data}
\label{sec:leaveoneout}
Table \ref{tab:leaveoneout} summarizes the findings for this experiment.
Like previously, ``no-enh'' and ``test-enh (*)'' serve as reference results.
In our case, we have four, namely, \textit{noise}, \textit{music}, \textit{babble}, \textit{chime3bg}.
In simulated data settings, usually, introduction of new noise classes in x-vector data leads to performance gains.
However, these augmentations can be harmful for real data, as established by the result rows which contain noise class name in first column.
They represent four \ac{SV} systems with x-vector data missing one noise class.
These results don't include enhancement and, thus, are comparable with ``no-enh'' system which has seen all noise classes.
We find omitting \textit{music} class in x-vector data gives best performance on \emph{BabyTrain}.
Similarly, omitting \textit{babble} and \textit{chime3bg} lead to performance better than ``no-enh''.
Speculating noise class which can hurt final performance is impossible a priori.
Therefore, speech enhancement is an appealing solution for improving robustness.

For all four \ac{SV} systems, we report the benefit of using our enhancement scheme.
``enh-seen'' and ``enh-unseen'' refer to cases when enhancement network training has or has not seen the noise class respectively.
Numbers in underline refer to best performance per \ac{SV} system.
Enhancement helped all four systems individually.
As expected, the enhancement system which has seen the missing noise class achieves the best performance (expect for \textit{noise}).
Importantly, this shows that enhancement helps even when a noise class is missing from x-vector training, regardless of whether it has seen that noise class itself or not.
However, ``test-enh (*)'' is worse than the best performance achieved in this ablation experiment, which reveals that current enhancement scheme is not strong enough to counter the degradation caused by harmful data augmentations.
This also highlights the scope in the improvement of the enhancement scheme.
Lastly, we note that omitting \textit{noise} (general environmental noises) brings degradation, suggesting the importance of complex environmental noises in training.
Thus, incorporating noise files from Voices2019~\cite{nandwana2019voices}, DCASE Challenge\footnote{\url{http://dcase.community/}}, and AudioSet~\cite{gemmeke2017audio} can be useful in our framework.

\vspace{-2mm}
\subsection{Handling reverberations}
\vspace{-1mm}
\label{sec:dereverb}
It is unclear if the \ac{DFL} based supervised enhancement scheme can work for the dereverberation task.
It is also unclear how much scope for dereverberation is in \emph{BabyTrain}.
In Table \ref{tab:dereverb}, we present results for several dereverberation schemes, some combined with denoising.
``WPE'' refers to Weighted-Prediction Error algorithm based pre-processing.
It gives minor improvement over ``no-enh''.
This suggests that dereverberation is either very challenging or has less scope in \emph{BabyTrain} in the first place.
``dereverb'' refers to \ac{DFL} system trained for only dereverberation, which gives worse performance than ``WPE'' suggesting \ac{DFL} scheme doesn't work for dereverberation out-of-the-box.
``WPE->denoise'' is the denoising system but with \ac{WPE} pre-processing.
It is minimally better than ``denoise''.
However, it is largely better than ``dereverb->denoise'', which refers to use of two \ac{DFL} systems trained (separately) for the two tasks respectively.
``denoise->dereverb'' (flipped version of ``dereverb->denoise'') does not lead to significant difference.

We now describe the joint training schemes.
``joint1stage'' refers to \ac{DFL} system trained for denoising and dereverberation (jointly) in one go.
Training pairs for it contain examples for denoising, dereverberation, and both.
Note that it is worse than ``WPE->denoise'' suggesting doing these two tasks in one-go is hard.
``joint2stage'' is an assisted modification of ``joint1stage''.
In addition to accepting reverberant and noisy signal input, it accepts another reverberant signal in the middle of the network and tries to minimise its deep feature loss as well.
This forces the network to first do denoising mimicking the standard signal model in signal processing.
This assisted scheme did not work, further solidifying our presumption that combining the tasks of denoising and dereverberation is very challenging.
Since our denoising network has seen few reverberant samples (from \textit{chime3bg}), we tried a double (disjoint) denoising scheme (``denoise->denoise'') and find it brings minimal improvement.
Results in this section suggest, finally, that the current \ac{DFL} scheme does not work for dereverberation and we suspect this problem is better solved through domain-adaptation methodology, as shown recently in \cite{nidadavolu2019low,nidadavolu2019unsupervised}.

\begin{table}[htbp]
    \centering
    \caption{Handling reverberations}
    \resizebox{0.47\textwidth}{!}{
        \begin{tabular}{|c|cccc|c|}
            \hline
            \textbf{EER} & \textbf{test>=30s} & \textbf{test>=15s} & \textbf{test>=5s} & \textbf{test>=0s} & \textbf{mean} \\
            \hline
                no-enh & 5.78 & 8.78 & 12.34 & 12.71 & 9.90\\
                \hline
                denoise (*) & \textbf{5.14} & \textbf{7.17} & 11.02 & 11.41 & 8.68\\
                \hline
                WPE & 5.94 & 8.65 & 12.20 & 12.58 & 9.84\\
                \hline
                WPE->denoise & 5.31 & 7.76 & 10.97 & 11.35 & 8.85\\
                \hline
                dereverb & 6.31 & 9.75 & 13.39 & 13.76 & 10.80\\
                \hline
                joint1stage & 5.63 & 8.33 & 11.55 & 11.93 & 9.36\\
                \hline
                joint2stage & 5.74 & 9.14 & 12.82 & 13.21 & 10.23\\
                \hline
                dereverb->denoise & 6.35 & 9.60 & 12.95 & 13.35 & 10.57\\
                \hline
                denoise->dereverb & 6.13 & 9.11 & 12.39 & 12.78 & 10.10\\
                \hline
                denoise->denoise & 5.26 & 7.59 & \textbf{10.73} & \textbf{11.11} & \textbf{8.67}\\
                \hline
                \multicolumn{6}{|c|}{}\\
                \hline
                \textbf{minDCF} & \textbf{test>=30s} & \textbf{test>=15s} & \textbf{test>=5s} & \textbf{test>=0s} & \textbf{mean} \\
                \hline
                no-enh & 0.255 & 0.386 & 0.492 & 0.499 & 0.408\\
                \hline
                denoise (*) & \textbf{0.204} & 0.333 & 0.441 & 0.448 & 0.357\\
                \hline
                WPE & 0.247 & 0.373 & 0.480 & 0.487 & 0.397\\
                \hline
                WPE->denoise & 0.206 & 0.330 & 0.438 & 0.445 & 0.355\\
                \hline
                dereverb & 0.242 & 0.391 & 0.498 & 0.504 & 0.409\\
                \hline
                joint1stage & 0.221 & 0.344 & 0.452 & 0.458 & 0.369\\
                \hline
                joint2stage & 0.249 & 0.393 & 0.501 & 0.508 & 0.413\\
                \hline
                dereverb->denoise & 0.249 & 0.394 & 0.499 & 0.506 & 0.412\\
                \hline
                denoise->dereverb & 0.244 & 0.386 & 0.492 & 0.499 & 0.405\\
                \hline
                denoise->denoise & 0.205 & \textbf{0.325} & \textbf{0.433} & \textbf{0.440} & \textbf{0.351}\\
                \hline
        \end{tabular}
    }
    \label{tab:dereverb}
    \vspace{-6mm}
\end{table}

\section{Conclusion}
Incorporating robustness in Speaker Verification is a challenging problem.
\emph{Data augmentation} usually handles this.
\emph{BabyTrain} is an appropriate dataset for this study due to its uncontrolled nature and emphasis on children's speech verification.
Since large data releases do not explicitly account for children speech, generalization of \ac{SV} systems to lower age group is an open question.
\emph{Deep feature loss} is a promising methodology which, in its current form, works along with \emph{data augmentation} in x-vector network.
It is shown to bring vast improvements.
Our experiments reveal that this test-time feature denoising approach is optimal when it utilizes all hidden activations of the auxiliary network excluding the final layer activations.
Search for best training data combination for enhancement and auxiliary network reveals that it is optimal to use top 50\% utterances of VoxCeleb according to their \ac{SNR}.
This satisfies the ideal requirements of clean, large, and naturalistic nature of data for training enhancement.
Experiments using different features for enhancement network shows it is best to use same 40-D \ac{LMFB} features as in the auxiliary and x-vector network.

An important inquiry into enhancing data other than test set reveals it is beneficial for the front-end (x-vector network) while harmful for the back-end (\ac{PLDA}).
To further investigate into the nature of enhanced signals, we hypothesize, and subsequently confirm, that they contain information complementary to the original signals.
We combine both signals in front-end and/or back-end to establish this.
The newly trained x-vector network with combined data turns out quite powerful as demonstrated by \textasciitilde40\% relative improvements over the baseline.
We also make a consistent observation that \ac{PLDA} is susceptible to enhancement processing.
The leave-one-out analysis solidifies the notion that \ac{DFL} enhancement is effective even when a noise class is missing from the training data of enhancement and/or x-vector network.
Importantly, we show the limitation of \emph{data augmentation} by demonstrating the degradation caused by including certain noise classes.
Surprisingly, they turn out to be common noise classes used in practice.
Finally, we design several dereverberation schemes combining \ac{WPE}, denoising, and dereverberation in either joint or disjoint fashion.
Extensive evaluation suggests ineffectiveness of \ac{DFL} enhancement for dereverberation while speculating domain-adaption as superior methodology.

We also speculate that findings of this work can vary with the choice of the x-vector network and the evaluation database.
Nevertheless, the analysis reported here provides further insight into the \emph{deep feature loss} based Speaker Verification and addresses its advantages, weaknesses, and extensions.

\clearpage

\bibliographystyle{IEEEbib}
\bibliography{Odyssey2020_BibEntries}

\end{document}